\documentclass[aip,a4paper,11pt,amsmath,amssymb]{revtex4-2}
\setcitestyle{numbers,square}
\usepackage{graphicx}
\usepackage{dcolumn}
\usepackage{subcaption} 
\usepackage{booktabs}
\usepackage{comment}
\usepackage{bm}

\begin{document}
\title[Nonlinear Power Absorption in CCRF Discharges]{Nonlinear Power Absorption in CCRF Discharges: Transition from Symmetric to Asymmetric Configurations}
\author{Katharina Noesges}
\email{noesges@aept.rub.de} 
\affiliation{Ruhr University Bochum, Chair of Applied Electrodynamics and Plasma Technology, D-44780 Bochum, Germany}
\author{Thomas Mussenbrock}
\email{mussenbrock@aept.rub.de}
\affiliation{Ruhr University Bochum, Chair of Applied Electrodynamics and Plasma Technology, D-44780 Bochum, Germany}
\date{\today}

\begin{abstract}
This work builds upon previous studies of nonlinear dynamics in low-pressure capacitively coupled radio-frequency discharges, focusing on the electron power absorption mechanism in discharges with various geometric asymmetries. We present a comprehensive investigation using a fully kinetic electrostatic 1d3v Particle-in-Cell/Monte Carlo collision simulation in spherical geometry. By systematically varying the inner electrode radius and the electrode gap distance, we analyze the influence of geometric asymmetry on key plasma properties, including electron density, power absorption, electron dynamics, and current characteristics. A central focus is placed on the cumulative power density as a diagnostic for energy deposition. In strongly asymmetric configurations, the cumulative electron power density exhibits distinct stepwise increases during sheath expansion, corresponding to the acceleration of successive electron beams. These nonlinear signatures are directly linked to the excitation of plasma series resonance and enhanced beam-driven power absorption. In contrast, more symmetric configurations display smoother, more symmetric cumulative power evolution, indicating balanced energy transfer at both sheaths and reduced nonlinearities. Time- and space-resolved diagnostics of cumulative power, current waveforms, and densities of energetic electrons reveal the critical role of asymmetry in shaping electron confinement and beam-driven power absorption. These findings demonstrate that the discharge geometry is actually an important design parameter which needs to taken into account during the design and construction phase of a reactor as it directly influences the plasma behavior with respect to energy deposition.
\end{abstract}

\maketitle

\section{\label{sec:introduction}Introduction}
Capacitively coupled radio-frequency (CCRF) discharges are widely used in industrial plasma applications, including thin-film deposition, plasma etching, and semiconductor processing \cite{Lieberman2005,Chabert2011,Bienholz2013}. The ability to control plasma parameters such as ion flux, electron density, and energy distribution is critical for optimizing these processes \cite{Kawamura1999,Makabe2015,Lee2005}. At low gas pressures, electron transport becomes nonlocal due to the long electron mean free path relative to the system dimensions, and sheath expansion plays a dominant role in determining power absorption mechanisms \cite{Schulze2011,Liu2017,Lee2019,Klich2021,Liu2022,Noesges2023}. 
In asymmetric discharges, these effects are more pronounced by geometric factors, such as the electrode area ratio, which directly influences the self-bias and the boundary sheaths and therefore, the power absorption\cite{Mussenbrock2006,Mussenbrock2006b,Mussenbrock2007,Mussenbrock2008,Czarnetzki2009,donko2009,czarnetzki2011,Wilczek2015,lafleur2016,Wilczek2016,Berger2018,Sharma2013,Bora2012,Bora2013,Bora2015}. Understanding the interplay between nonlocal electron dynamics, sheath motion, and nonlinear resonances is crucial for predicting plasma behavior in these systems. A key phenomenon in low-pressure CCRF discharges is the development of a DC self-bias, which arises from unequal electrode areas and results in an asymmetric sheath voltage distribution. This leads to the self-excitiation of the plasma series resonance (PSR) due to nonlinear interaction between the plasma bulk and the boundary sheath \cite{Klick1996,Czarnetzki2006,Mussenbrock2006,Yamazawa2009,Schungel2011,Lieberman2015,Wen2017,wen2017a,Oberberg2018,Oberberg2019}. These oscillations lead to the formation of electron beams that propagate through the plasma bulk. The interaction of these beams with the opposing sheath determines the electron dynamics and, therefore, the plasma parameters \cite{Liu2011,Liu2012,Liu2012b}. To investigate these phenomena, we employ a 1d3v electrostatic Particle-in-Cell/Monte Carlo collision (PIC/MCC) simulation based on the \textit{yapic} code \cite{Turner2013}. The code has been extended to support spherical coordinates, allowing us to simulate discharges with tunable asymmetry by varying the inner electrode radius. Using the same parameters as those used in a prior PIC code benchmark \cite{Turner2013}, we analyze the influence of geometric asymmetry on electron power absorption, sheath behavior, and nonlinear resonance phenomena. This study focuses on low-pressure conditions, where collisional damping is minimal and nonlinear effects dominate. The results highlight how geometric factors affect the excitation of nonlinear resonances, the formation of electron beams, and overall power coupling efficiency. In particular, asymmetric discharges are found to support nonlinear oscillations and more efficient beam-driven power absorption compared to quasi-symmetric setups.

The paper is structured as follows: Section 2 outlines the simulation setup, including the details of the geometric configurations and the implementation of the spherical PIC/MCC model. Section 3 presents the results, starting with a comparison between Cartesian 1d and spherical 1d geometries, followed by a detailed analysis of how geometric asymmetry affects power absorption, electron dynamics, and nonlinear excitation. Finally, Section 4 summarizes the key findings and discusses their implications for optimizing CCRF discharges under asymmetric conditions.

\section{Model and Simulation Approach}

\begin{figure}[h!]
    \centering
    \includegraphics[width=\textwidth]{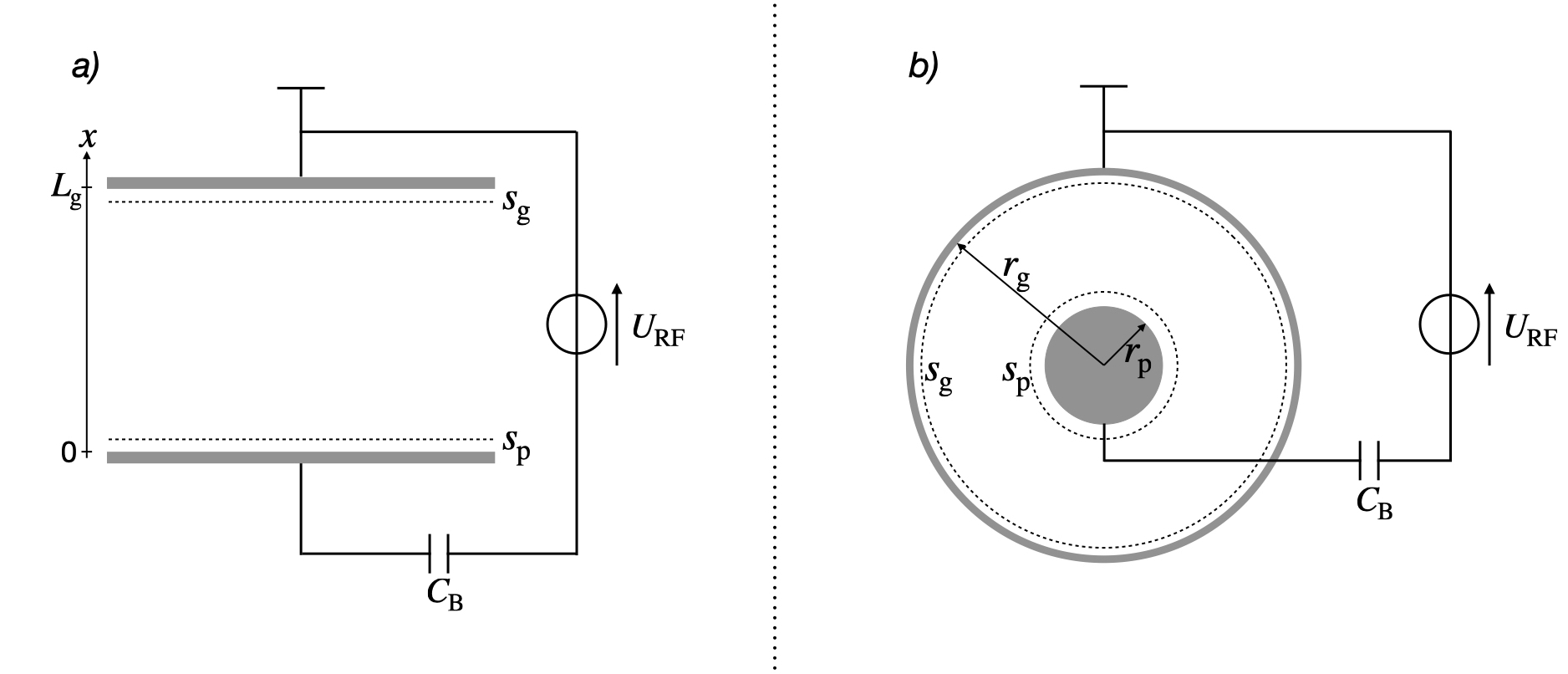}
    \caption{Capacitive coupled radio frequency discharge using a Cartesian 1d (a) and spherical 1d (b) geometry}
    \label{fig:Geometry}
\end{figure}

Geometrically symmetric and asymmetric CCRF discharges are investigated using the  benchmarked electrostatic 1d3v PIC/MCC code \textit{yapic} \cite{Turner2013}. For the symmetric case, the simulation domain is confined between two plane-parallel electrodes, where the plasma dynamics are resolved along the \(x\)-direction (c.f. fig.\ref{fig:Geometry}(a)). The bottom electrode is powered, while the top electrode is grounded. In this case, the resulting electrode area ratio between the grounded and powered electrodes is \( A_{\text{g}} / A_{\text{p}} = 1.0 \), meaning that the area of the powered electrode is equally in size to the grounded electrode.  In the asymmetric configuration (c.f. fig.\ref{fig:Geometry}(b)), the discharge is confined between two concentric spheres located at \(r = r_{\rm{p}} \) and \(r = r_{\rm{g}} \). The inner spherical shell, which serves as the powered electrode, is connected to voltage source via a blocking capacitor. The outer spherical shell represents the grounded electrode. Due to the spherical symmetry of the system, only the radial variation of the plasma is considered. The resulting asymmetry factor \( \epsilon \) is defined as the ratio of the grounded to the powered electrode areas in the spherical geometry. It depends on the inner radius \( r_{\rm{p}} \), the outer radius \( r_{\rm{g}} \), and the electrode gap distance \( L_{\rm{g}} \), and is given by
\begin{equation}
\epsilon = \frac{A_{\text{g}}}{A_{\text{p}}} = \frac{r_{\rm{g}}^2}{r_{\rm{p}}^2} = \left(\frac{r_{\rm{p}} + L_{\rm{g}}}{r_{\rm{p}}}\right)^2.
\label{eq:asymmetry_factor}
\end{equation}
The schematic in Figure~\ref{fig:Geometry} indicates that the spherical geometry leads to different sheath widths in front of the powered and grounded electrodes (\( s_{\mathrm{p}} \) and \( s_{\mathrm{g}} \)), which in turn influences the discharge behavior. The details of the spherical implementation are described in great detail by Verboncoeur et al. \cite{Verboncoeur1991}. The voltage source for both geometries provides a sinusoidal waveform:\[U(t) = V_0 \sin(2 \pi f_{\rm{RF}} t),\] where \( f_{\rm{RF}} \) is the radio frequency (RF). In all configurations, a voltage amplitude \( V_0 = 500\,\mathrm{V} \), a pressure \( p = 1\,\mathrm{Pa} \), and an Argon background gas at a temperature \( T_{\rm{g}} = 300\,\mathrm{K} \) are used. The collision mechanisms include three electron-neutral processes (elastic scattering, excitation, and ionization). For Helium the cross section set of Biagi-v7.1 is used and for Argon the Phelps database is used both are obtained via the website of the LXCat project \cite{Phelps1994,Pitchford2017,Pancheshnyi2012,Alves2014,lxcat}. For the heavy particles two ion-neutral processes (isotropic and backward elastic scattering). Cross-section data for these processes are sourced from the Phelps database \cite{Phelps1994, lxcat}. Plasma surface interactions, such as particle reflection and secondary electron emission, are neglected to simplify comparisons between the two geometries. For both geometries, Cartesian and spherical, an equidistant grid is employed.
The spatial resolution is given by \( \Delta x = L_{\text{g}} / N_{\text{cells}} \) or \( \Delta r = L_{\text{g}} / N_{\text{cells}} \), where \( N_{\text{cells}} \) is the number of grid cells and is adjusted so that the Debye length $\lambda_{\rm{d}}$ \cite{Birdsall1991b,Donko2021,Turner2013} is resolved. The time step is set as \( \Delta t = 1 / (f_{\rm{RF}} \cdot N_{\rm{tspc}}) \), where \( N_{\rm{tspc}} \) is the number of time steps per RF cycle  and is chosen to satisfy the requirements
regarding the electron plasma frequency \cite{Birdsall1991b,Donko2021,Turner2013,Wilczek2020} and to ensure that the Courant-Friedrichs-Lewy criterion for numerical stability \cite{Birdall1991,Birdsall1991b,Hockney1988} is always fulfilled. Approximately \( 1.0 \times 10^5 \) superparticles per electron and ion species are targeted after convergence is achieved.

\section{Results}

\subsection*{Comparison of the geometries}

In this subsection, the spherical PIC code is compared with the benchmarked Cartesian version of the \textit{yapic} code to lay the basis for investigating the nonlinear phenomena in asymmetric discharges in the following sections.

\begin{figure}[h!]
  \centering
    \includegraphics[width=\linewidth]{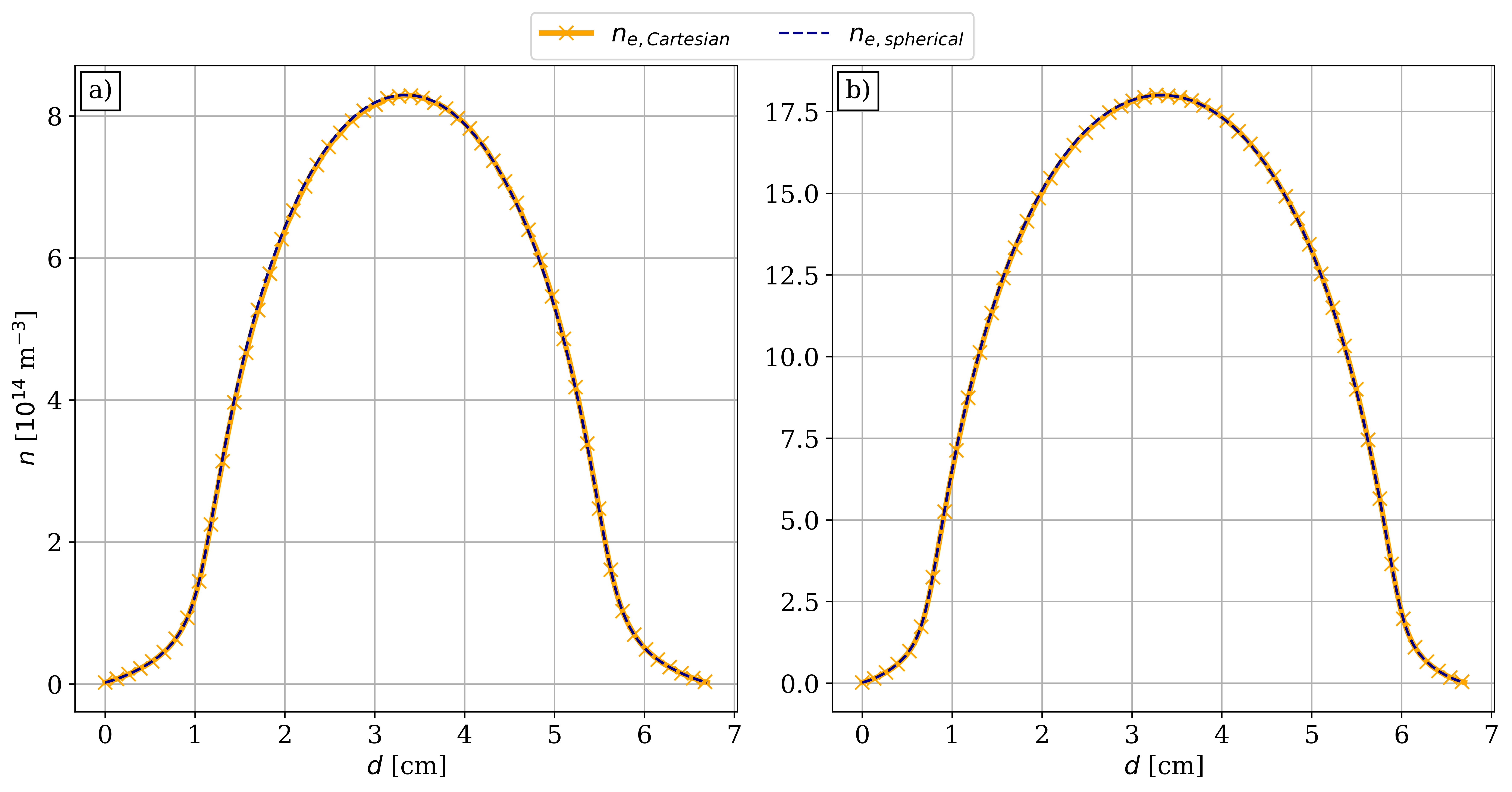}
\caption{Comparison of the averaged electron densities for the Cartesian geometry (orange curve) and the spherical geometry (blue dashed curve) for benchmark case 2 (a) and 3 (b) in Helium (c.f. \cite{Turner2013}).}
  \label{fig:CompXvsR}
\end{figure}

Figure~\ref{fig:CompXvsR} illustrates the time-averaged electron density as a function of the distance from the driven electrode. The Figure compares the density profiles for benchmark condition 2 and 3 in both Cartesian and spherical geometries. The benchmark cases are described in detail by Turner \textit{et al.}~\cite{Turner2013}. To achieve geometric symmetry with the spherical configuration, the inner radius is set to \( r_{\rm{p}} = 100.0 \, \text{cm} \). This results in an almost exact agreement with the Cartesian results for both benchmark condition 2 and 3, indicating that the spherical configuration is suitable for describing geometrically symmetric setups. We refer to this as "quasi-symmetric". Based on this agreement, the spherical version of the code is now applied to investigate the transition from symmetric to asymmetric configurations.

\subsection*{Influence of the spherical geometry on the discharge dynamics}
In this section, the asymmetry factor $\epsilon$ for the spherical geometry is varied to investigate its influence on the discharge behavior in Argon.

 \begin{figure}[h]
     \centering
     \begin{subfigure}{0.48\textwidth}
         \centering
         \includegraphics[width=\linewidth]{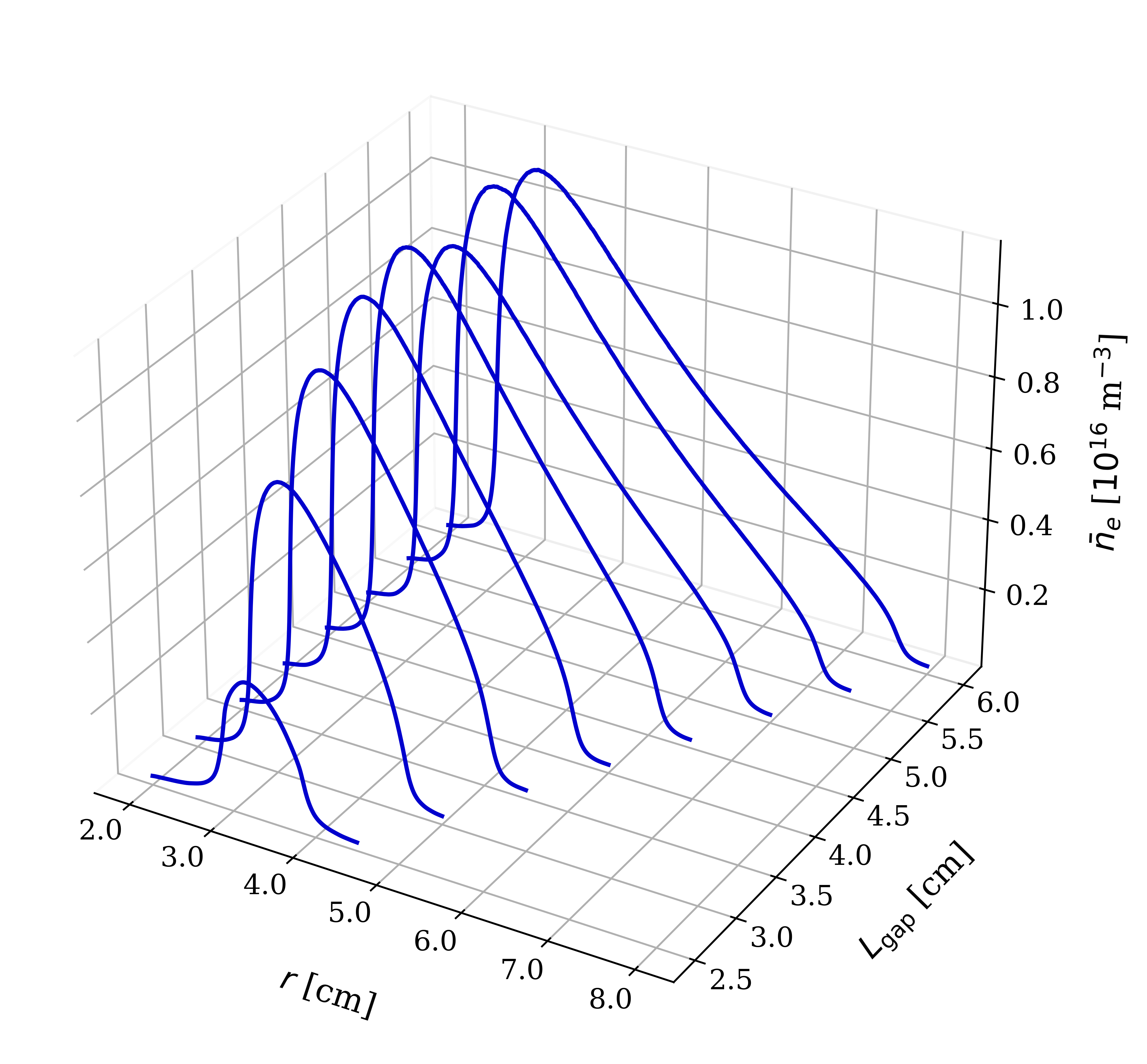}
         \caption{$r_{\rm{p}}\,=\,2.0\,\rm{cm}$}
         \label{fig:sub1}
     \end{subfigure}
     \hfill
    \centering
     \begin{subfigure}{0.48\textwidth}
         \centering
         \includegraphics[width=\linewidth]{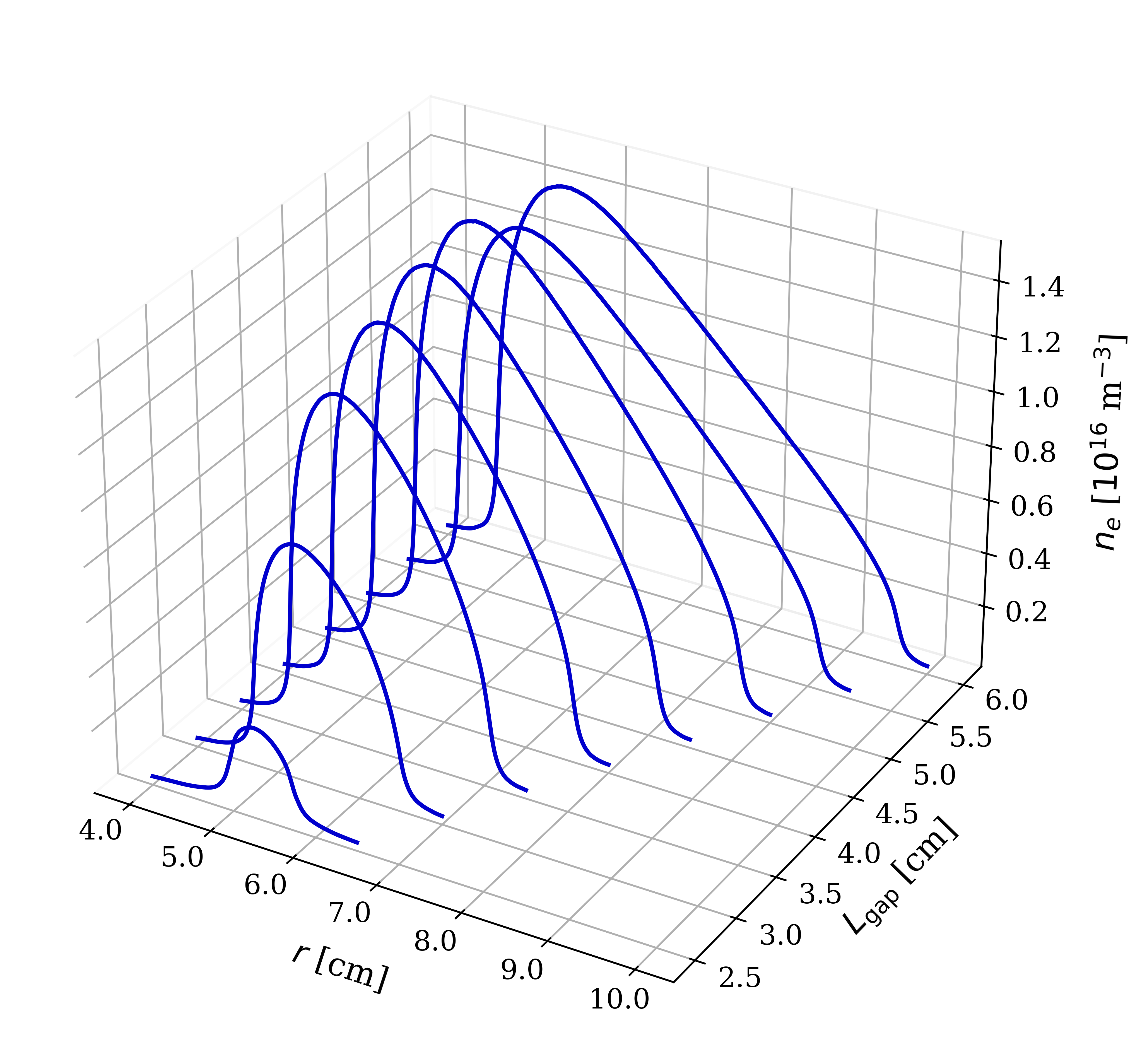}
         \caption{$r_{\rm{p}}\,=\,4.0\,\rm{cm}$}
         \label{fig:sub2}
     \end{subfigure}
     \hfill
     \begin{subfigure}{0.48\textwidth}
         \centering
         \includegraphics[width=\linewidth]{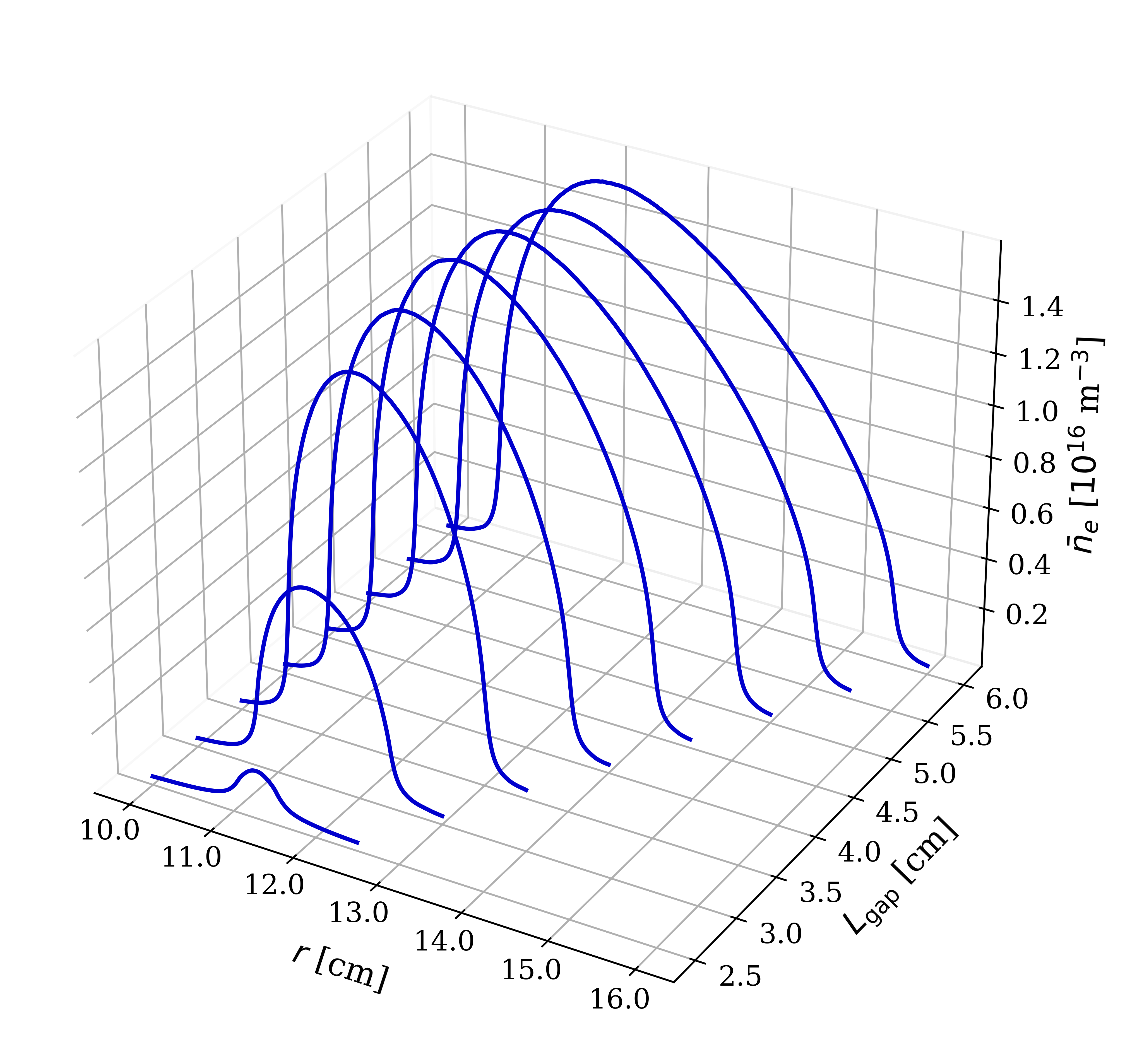}
         \caption{$r_{\rm{p}}\,=\,10.0\,\rm{cm}$}
         \label{fig:sub3}
     \end{subfigure}
          \hfill
     \begin{subfigure}{0.48\textwidth}
         \centering
         \includegraphics[width=\linewidth]{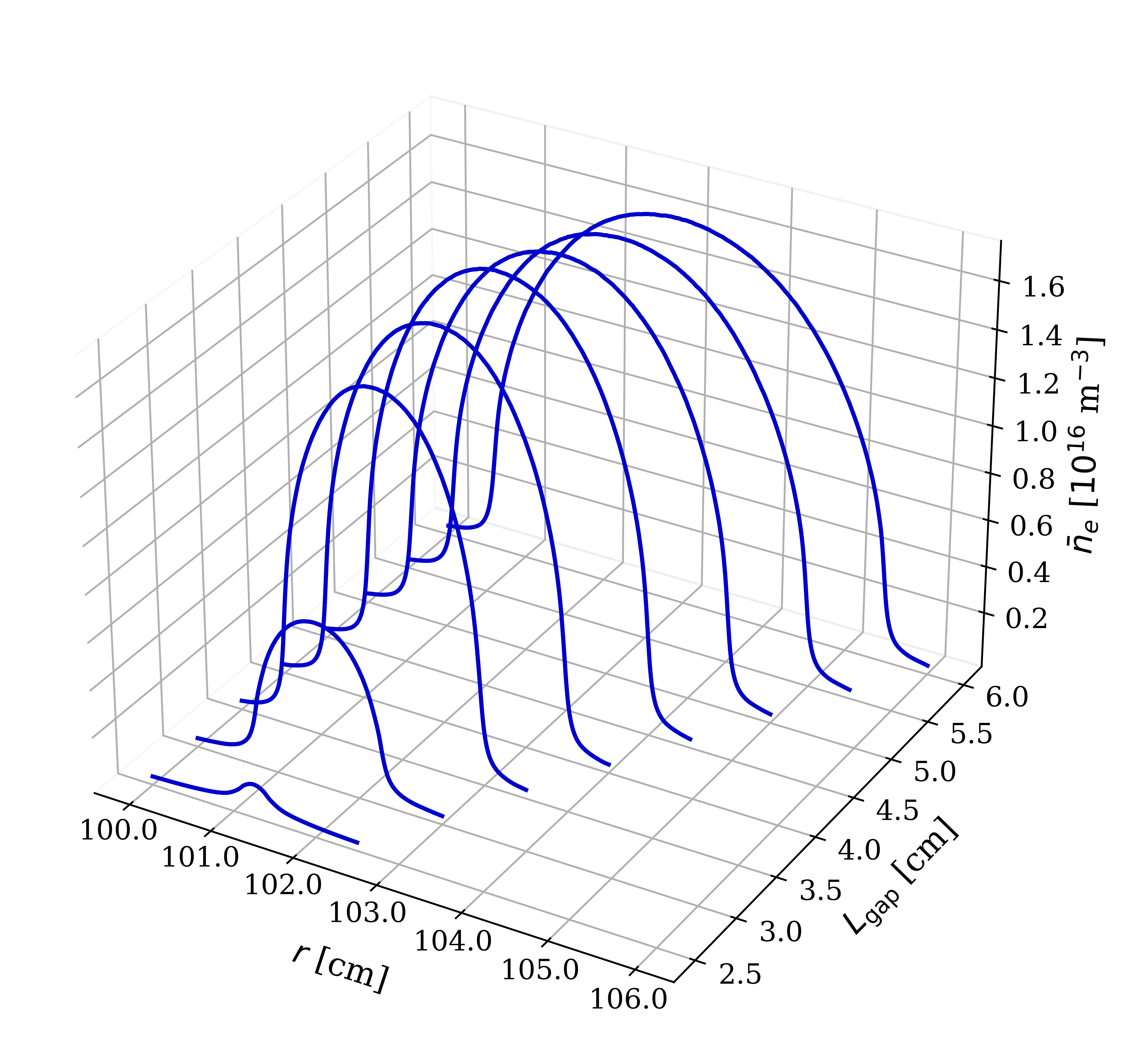}
         \caption{$r_{\rm{p}}\,=\,100.0\,\rm{cm}$}
         \label{fig:sub4}
     \end{subfigure}
     \caption{Comparison of the time averaged electron densities for different inner radii and gap distances. Discharge parameters: $V_{0} = 500\,\rm{V},\ f=27.12\,\rm{MHz},\ p = 1.0\,\rm{Pa},\ 100\, \% \,\rm{Ar}$}
     \label{fig:AveDensitiesAr1Pa}
 \end{figure}

\noindent Figure~\ref{fig:AveDensitiesAr1Pa} illustrates the time-averaged electron density as a function of the distance from the driven electrode for various inner radii and gap distances. The four subfigures represent inner radii \( r_{\rm{p}} = 2.0\, \rm{cm}, 4.0\, \rm{cm}, 10.0\, \rm{cm}, \) and \( 100.0\, \rm{cm} \), highlighting how the electron density profile varies with electrode gap size and geometric asymmetry. For a smaller inner radius of \( r_{\rm{p}} = 2.0\, \rm{cm} \) (c.f. fig. \ref{fig:AveDensitiesAr1Pa}(a)), the maximum electron density increases substantially, from \( n_{\mathrm{e}} = 3.5 \cdot 10^{15}\ \mathrm{m}^{-3} \) at a gap distance of \( L_{\mathrm{g}} = 2.5\, \rm{cm} \) to \( n_{\mathrm{e}} = 1.1 \cdot 10^{16}\ \mathrm{m}^{-3} \) at \( L_{\mathrm{g}} = 6.0\, \rm{cm} \). The electron density profile becomes increasingly asymmetric with larger gap distances, as the peak density shifts toward the smaller powered electrode ( \( r = r_{\rm{p}} \)). This corresponds to an increase in the asymmetry factor from \( \epsilon = 5.1 \) to \( \epsilon = 16.0 \). When the inner radius is increased to \( r_{\rm{p}} = 4.0\, \rm{cm} \) (c.f. fig. \ref{fig:AveDensitiesAr1Pa}(b)), a similar trend is observed, with the maximum electron density reaching \( n_{\mathrm{e}} = 1.3 \cdot 10^{16}\ \mathrm{m}^{-3} \) at \( L_{\mathrm{g}} = 6.0\, \rm{cm} \). Overall, less asymmetric electron density profiles are observed. For an inner radius of \( r_{\rm{p}} = 10.0\, \rm{cm} \) (c.f. fig. \ref{fig:AveDensitiesAr1Pa}(c)), the density profiles exhibit broader distributions and higher peak densities, except for small electrode gaps (\( L_{\mathrm{g}} = 2.5\, \rm{cm} - 3.0\, \rm{cm} \)). This behavior can be attributed to different power absorption mechanisms driven by different dynamics of the sheath in front of the driven and the grounded electrode, which will be discussed in more detail later in this paper. At the largest inner radius (\( r_{\rm{p}} = 100.0\, \rm{cm} \)) (c.f. fig. \ref{fig:AveDensitiesAr1Pa}(d)), the discharge becomes quasi-symmetric, with the peak electron density centered between the inner and outer electrodes. For this configuration, the highest peak density is observed for \( L_{\mathrm{g}} = 4.5\, \rm{cm} \) at  \( n_{\mathrm{e}} = 1.75 \cdot 10^{16}\ \mathrm{m}^{-3} \), where the maximum of bounce resonance heating is assumed \cite{Liu2011,Liu2012}. For small gap sizes, a reversed trend is observed across all inner radii when compared to larger gap distances (\( L_{\mathrm{g}} = 3.5 - 6.0\,\mathrm{cm} \)). In this regime, the discharge with the highest asymmetry (associated with the smallest inner radius) produces the highest electron density. This behavior can be attributed to distinct electron dynamics that arise under strongly asymmetric conditions, which will be examined in more detail in the following sections.

\begin{figure}[h!]
    \centering
    \includegraphics[width=\textwidth]{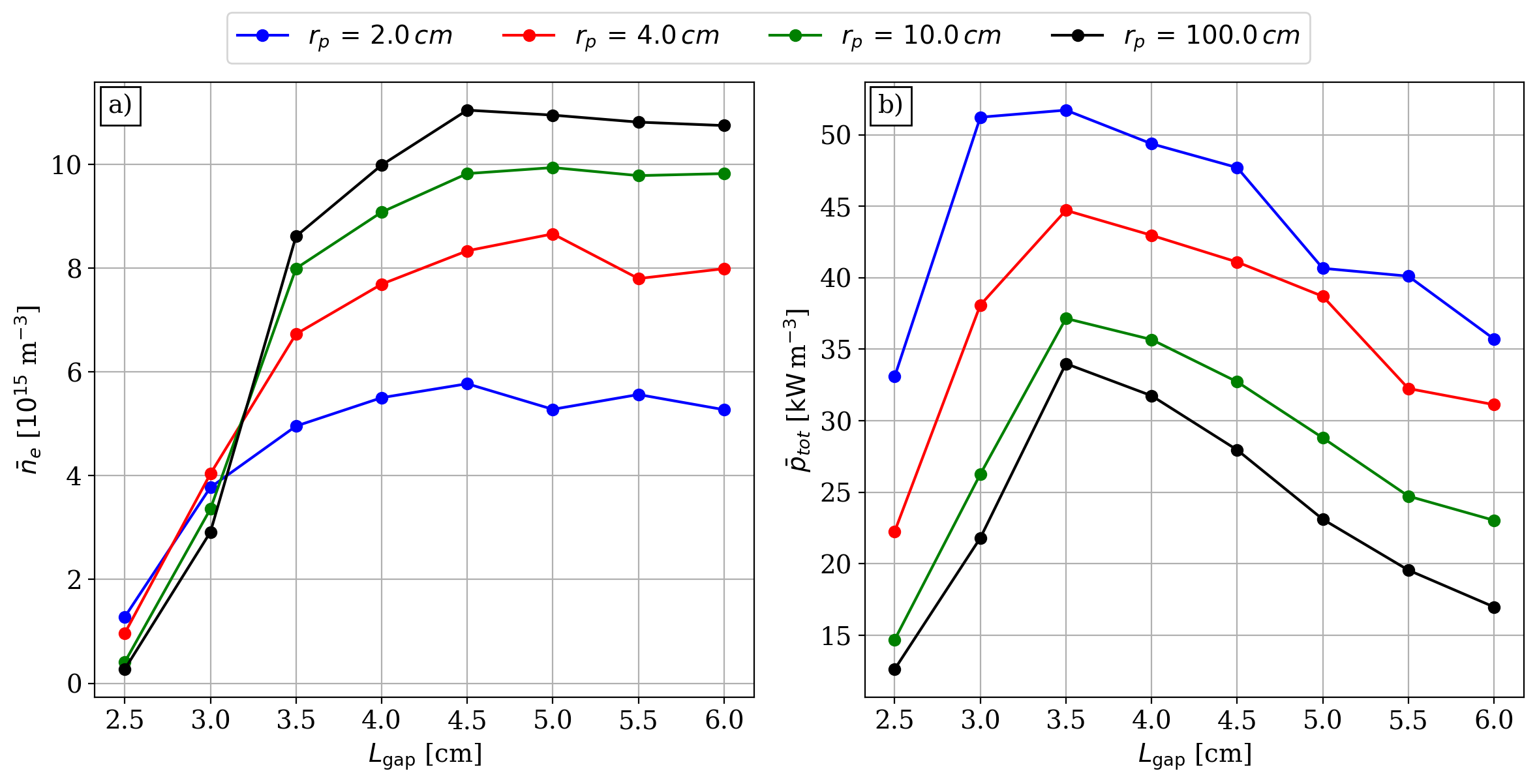}
    \caption{Comparison of the time- and space averaged densities (a) and powers (b) for different inner radii \( r_{\rm{p}} = 2.0\,\rm{cm},\ 4.0\,\rm{cm},\ 10.0\,\rm{cm}, \),\( 100.0\,\rm{cm} \). Discharge parameters: $V_{0} = 500\,\rm{V},\ f=27.12\,\rm{MHz},\ p = 1.0\,\rm{Pa},\ 100\, \% \,\rm{Ar}$}
    \label{fig:DenPow}
\end{figure}

Figure~\ref{fig:DenPow} presents the time- and space-averaged electron densities and power densities for various inner radii and gap distances. The blue, red, green, and black curves correspond to inner radii of \( r_{\rm{p}} = 2.0\,\rm{cm},\ 4.0\,\rm{cm},\ 10.0\,\rm{cm}, \) and \( 100.0\,\rm{cm} \), respectively. The trends observed in Figure~\ref{fig:DenPow}(a) align with those in Figure~\ref{fig:AveDensitiesAr1Pa}, confirming the influence of geometric asymmetry and gap size on the plasma characteristics. At small gap distances, the electron density is generally low, but it increases with increasing gap size and eventually saturates, consistent with trends reported in symmetric discharges~\cite{Noesges2023}. Figure~\ref{fig:DenPow}(b) shows the corresponding power densities, which increase with gap size and reach a maximum at \( L_{\mathrm{g}} = 3.5\,\rm{cm} \), indicating optimal power coupling. Beyond this point, the power density decreases with further increase in the gap distance. Notably, for small gaps (\( L_{\mathrm{g}} = 2.5 - 3.0\,\rm{cm} \)), the highest electron densities are obtained in the most asymmetric configuration (\( r_{\rm{p}} = 2.0\,\rm{cm} \)), highlighting the effectiveness of the power absorption at small inner radii. However, for larger gap distances (\( L_{\mathrm{g}} = 3.5 - 6.0\,\rm{cm} \)), this trend reverses: discharges with larger inner radii yield higher electron densities. For instance, at \( L_{\mathrm{g}} = 6.0\,\rm{cm} \), the quasi-symmetric configuration with \( r_{\rm{p}} = 100.0\,\rm{cm} \) achieves a total power density of \( \bar{p}_{\mathrm{tot}} = 17\,\mathrm{kW/m^3} \) and a corresponding electron density of \( n_{\mathrm{e}} = 1.1 \times 10^{16}\,\mathrm{m}^{-3} \). In contrast, for \( r_{\rm{p}} = 2.0\,\rm{cm} \) a higher power density of \( \bar{p}_{\mathrm{tot}} = 36\,\mathrm{kW/m^3} \) is calculated, but a much lower electron density of \( n_{\mathrm{e}} = 5.3 \times 10^{15}\,\mathrm{m}^{-3} \). These observations highlight the opposing trends between power coupling and electron density. While strong asymmetry favors high power deposition, it does not guarantee high plasma densities. Instead, symmetric discharges become more efficient at sustaining high electron densities as the gap increases.

\begin{figure}[h!]
    \centering
    \includegraphics[width=\textwidth]{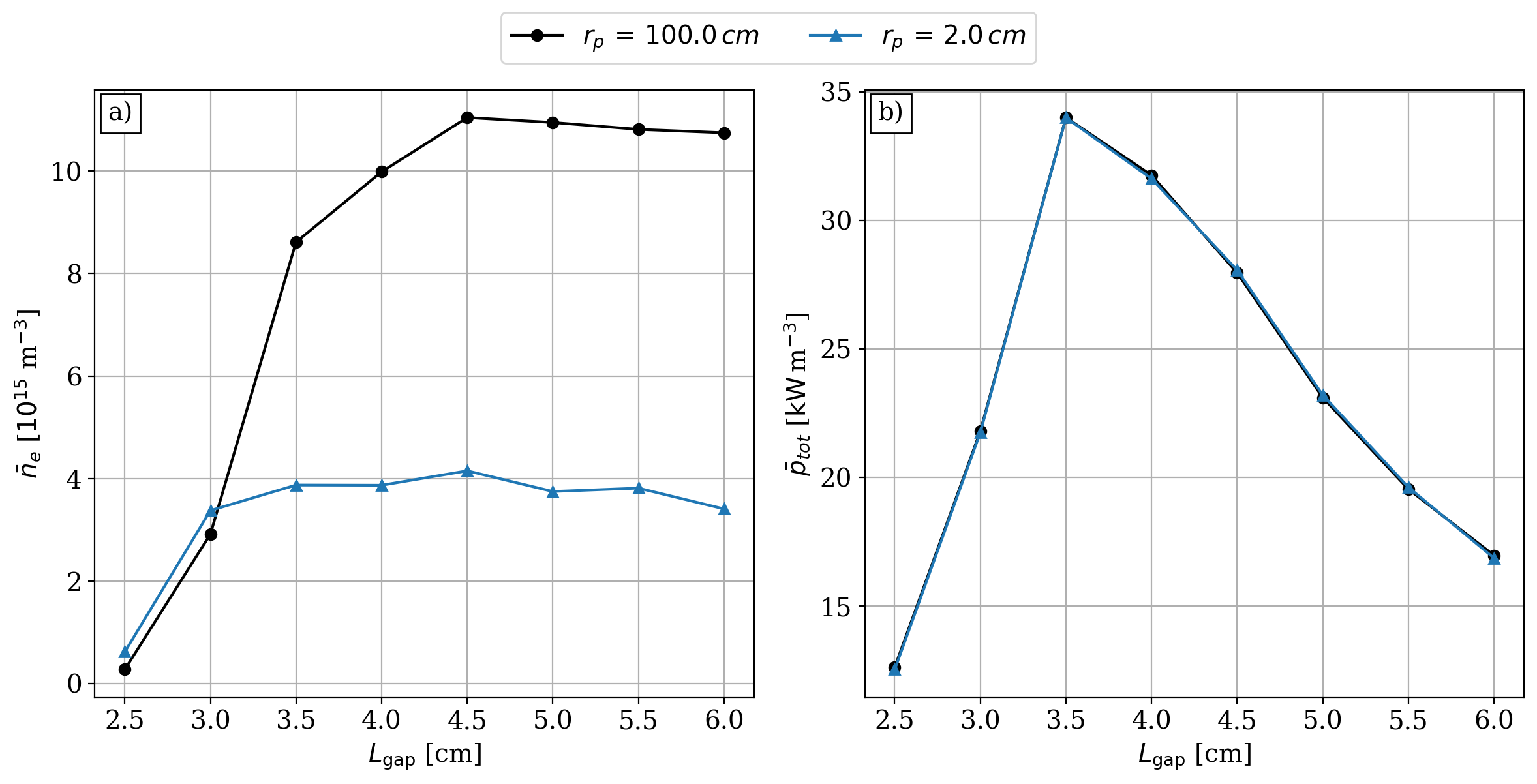}
    \caption{Comparison of the averaged electron densities at different geometric configurations (a) with the identical power coupling (b). Discharge parameters: $V_{0} = 500\,\rm{V},\ f=27.12\,\rm{MHz},\ p = 1.0\,\rm{Pa},\ 100\, \% \,\rm{Ar}$}
    \label{fig:DenPowEqualized}
\end{figure}

To ensure that the observed trends in electron density are not merely a consequence of different power coupling efficiencies across configurations, we now present a complementary analysis where the total power density to the system is adjusted to match that of the quasi-symmetric reference case. This approach allows a direct comparison of the intrinsic effectiveness of each discharge geometry in converting absorbed power into electron density. Figure~\ref{fig:DenPowEqualized} shows the time- and space-averaged electron density for \( r_{\rm{p}} = 100.0\,\mathrm{cm} \) and \( r_{\rm{p}} = 2.0\,\mathrm{cm} \), where the dissipated power density is identical in both cases. The resulting blue curve reveals trends that are consistent with the original observations in Figure~\ref{fig:DenPow}(a). While the absolute values of the electron density are reduced across all gap distances due to the lower deposited power, the characteristic behavior remains unchanged. At small gap distances (\( L_{\mathrm{g}} = 2.5 - 3.0\,\mathrm{cm} \)), the asymmetric configuration still produces higher electron densities compared to the symmetric case (black curve), confirming the efficiency of power absorption mechanisms under strong asymmetry. However, as the gap distance increases, the electron density for the symmetric configuration rises sharply, whereas the electron density for \( r_{\rm{p}} = 2.0\,\mathrm{cm} \) saturates more gradually. This supports the conclusion that the observed trends are not merely artifacts of differing power input but are intrinsically linked to the underlying discharge physics and geometric asymmetry.

\begin{table}[h!]
\centering
\begin{tabular}{lcr}
  \toprule
  & $L_{\rm{g}} = 3.0\, \rm{cm}\ \ $   &$ \ \ \ \  L_{\rm{g}} = 6.0\, \rm{cm}$\\
  \midrule
$r_{\rm{p}} = 2.0\, \rm{cm}$  & 6.25  & 16.00 \\
$r_{\rm{p}} = 4.0\, \rm{cm}$  & 3.06  & 6.25 \\
$r_{\rm{p}} = 10.0\, \rm{cm}$ & 1.69  & 2.56 \\
$r_{\rm{p}} = 100.0\, \rm{cm}$& 1.06  & 1.12 \\
  \bottomrule
\end{tabular}
\caption{Asymmetry factor $\epsilon$ for different inner radii and gap distances calculated using equation \eqref{eq:asymmetry_factor}.}
\label{AsymmetryFactors}
\end{table}

\noindent To further investigate these effects, the time-resolved cumulative power absorption is analyzed for both a small-gap case (\( L_{\mathrm{g}} = 3.0\,\mathrm{cm} \)) and a large-gap case (\( L_{\mathrm{g}} = 6.0\,\mathrm{cm} \)) across the full range of inner radii. This analysis provides deeper insight into the temporal evolution of power deposition and its dependence on geometric asymmetry. The corresponding asymmetry factors, calculated using equation \eqref{eq:asymmetry_factor}, are listed in Table~\ref{AsymmetryFactors}.

\begin{figure}[h!]
    \centering
    \includegraphics[width=\textwidth]{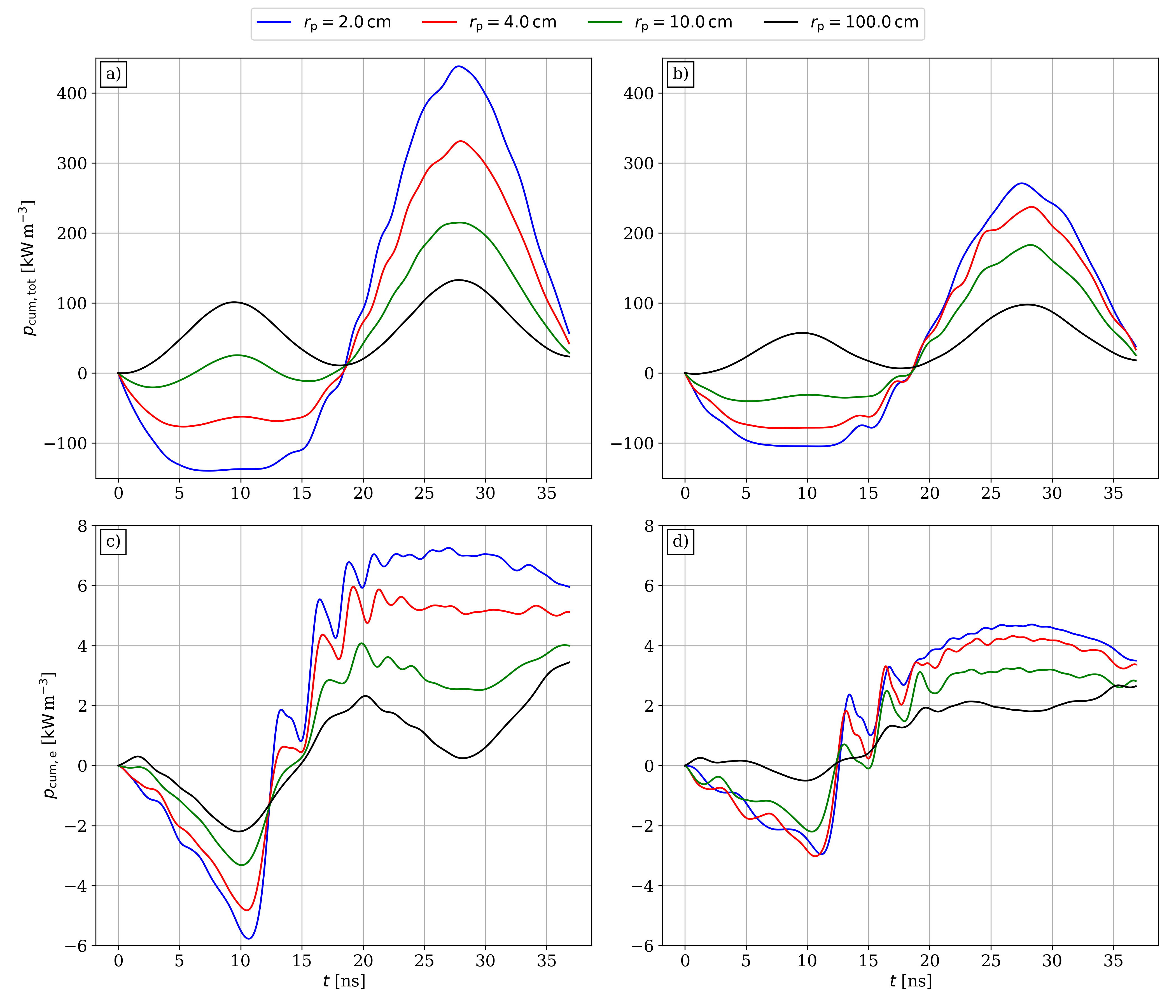}
    \caption{Comparison of the total cumulative power density (first row) and the cumulative electron power density (second row) for \( L_{\mathrm{g}} = 3.0\ \rm{cm} \) (left column) and \( L_{\mathrm{g}} = 6.0\ \rm{cm} \) (right column). Discharge parameters: $V_{0} = 500\,\rm{V},\ f=27.12\,\rm{MHz},\ p = 1.0\,\rm{Pa},\ 100\, \% \,\rm{Ar}$}
    \label{fig:CummPower}
\end{figure}

Figure~\ref{fig:CummPower} presents the results of this comparison. The first row displays the total cumulative power density, while the second row shows the cumulative electron power density. The cumulated power density is calculated using:

\begin{align}
p_{\text{cum}}(t) = \frac{1}{T_{\mathrm{RF}}} \int_{0}^{t} \bar{p}(t') \, dt',
\end{align}

\noindent where \( \bar{p}(t') \) denotes the instantaneous total or electron power density spatially averaged over the electrode gap. The blue, red, green, and black curves correspond to inner radii of \( r_{\rm{p}} = 2.0\ \rm{cm},\ r_{\rm{p}} = 4.0\ \rm{cm},\ r_{\rm{p}} = 10.0\ \rm{cm}, \) and \( r_{\rm{p}} = 100.0\ \rm{cm} \), respectively, covering the transition from highly asymmetric to quasi symmetric configurations. At the beginning of the RF period, all curves start at zero, indicating no cumulative power absorption initially. Over time, the curves diverge, reflecting differences in power absorption dynamics based on the inner radius and gap distance. In Figure~\ref{fig:CummPower}(a), the total cumulative power density is shown for \( L_{\mathrm{g}} = 3.0\,\rm{cm} \). For the smallest inner radius (\( r_{\rm{p}} = 2.0\,\rm{cm} \)), the power density initially decreases during the first half of the RF cycle, indicating a net power loss. As the sheath at the powered electrode expands during the second half, the system rapidly gains power. This results in a stepwise increase in cumulative power density, characteristic of the system's nonlinear dynamics. By the end of the RF period, the cumulative power aligns with the time- and space-averaged power density. As the inner radius increases to \( r_{\rm{p}} = 4.0\,\rm{cm} \), a similar trend is observed, but with reduced peak-to-peak amplitude. Additionally, at larger radii (\( r_{\rm{p}} = 10.0\,\mathrm{cm} \) and \( 100.0\,\mathrm{cm} \)), the power absorption transitions to positive values at approximately \( t \approx 10\,\mathrm{ns} \).
For these larger inner radii, the cumulative power density evolves more symmetrically due to increasing similarity between the sheath electric fields and current densities at both electrodes. As a result, two distinct power gain maxima appear, indicating more balanced energy coupling. Furthermore, the discrete steps in the power density diminish as the discharge geometry becomes more symmetric and nonlinear effects are reduced. Figure~\ref{fig:CummPower}(b) shows the corresponding results for a larger gap distance of \( L_{\mathrm{g}} = 6.0\,\rm{cm} \). As indicated by the asymmetry factors in Table~\ref{AsymmetryFactors}, this configuration exhibits the highest degree of asymmetry. Consequently, only the curve corresponding to the largest inner radius (\( r_{\rm{p}} = 100.0\,\rm{cm} \)) displays net power gain during both halves of the RF cycle, reflecting quasi symmetric sheath dynamics. For smaller radii, the first half of the cycle is dominated by power loss, as the grounded sheath contributes little to the net energy transfer. The appearance of distinct steps in the curves highlights the increased importance of nonlinear interactions, particularly the formation of multiple electron beams during the sheath expansion phase. It is important to note that the total cumulative power density primarily accounts for the displacement current in the sheath electric field. To understand the energy transfer to the electron population, the cumulative electron power density is investigated.

The blue curve in Figure~\ref{fig:CummPower}(c) illustrates the temporal evolution  of the cumulative electron power density. Initially, until \( t \approx 10\, \mathrm{ns} \), the electrons lose energy as the sheath at the grounded electrode collapses. Once the sheath at the powered electrode begins to expand, a rapid power gain is observed, resulting in a local maximum around \( t \approx 12\,\mathrm{ns} \), which corresponds to the first electron beam being accelerated into the plasma bulk. Subsequent maxima are associated with additional electron beams formed during sheath expansion. Analyzing the electron dynamics and the interaction between both sheaths provides insight into how the discharge geometry affects power absorption — an aspect explored in more detail later. When increasing the inner radius to \( r_{\rm{p}} = 10.0\,\mathrm{cm} \) and \( r_{\rm{p}} = 100.0\,\mathrm{cm} \), the pronounced local maxima vanish, and the cumulative electron power density increases more smoothly, reflecting more balanced and symmetric sheath behavior. The black curve clearly demonstrates how electrons gain and lose energy at both sheaths. Between \(t \approx 0 - 10\,\mathrm{ns} \), the collapse of the sheath at the powered electrode leads to negative cumulative power. From \(t \approx 10 - 20\,\mathrm{ns} \), a positive cumulative power density is observed as electrons are accelerated by the sheath at the powered electrode. From \( t \approx 20 - 30\,\mathrm{ns} \), the collapse of the sheath at the grounded electrode causes a reduction in the cumulative power, though the value does not turn negative. After \( t \approx 27\,\mathrm{ns} \), when the sheath at the grounded electrode begins to expand again, the cumulative power increases, reflecting the interplay between the two equal sheaths. The blue curve in Figure~\ref{fig:CummPower}(d) shows similar behavior to that in Figure~\ref{fig:CummPower}(c), but with more pronounced first and second local maxima. In contrast, the third and fourth local maxima that were previously visible are now suppressed at larger gap distances. The cumulative power density increases up to \( t \approx 27\,\mathrm{ns} \), followed by a decrease due to the collapse of the sheath at the powered electrode. The contribution of the grounded sheath to the overall power absorption is negligible in this configuration. This is due to the strong geometric asymmetry (\( \epsilon = 16.0 \)), which results in electron power gain and loss occurring primarily at the powered electrode. For the largest inner radius, the cumulative power density exhibits a much smoother temporal evolution compared to black curve shown in Figure~\ref{fig:CummPower}(c). In particular, the decrease in power density after \( t \approx 20\,\mathrm{ns} \) is barely visible. This smoothing effect can be attributed to impingement phases of the electron beams on the opposing sheath. 

\begin{figure}[h!]
    \centering
    \includegraphics[width=\textwidth]{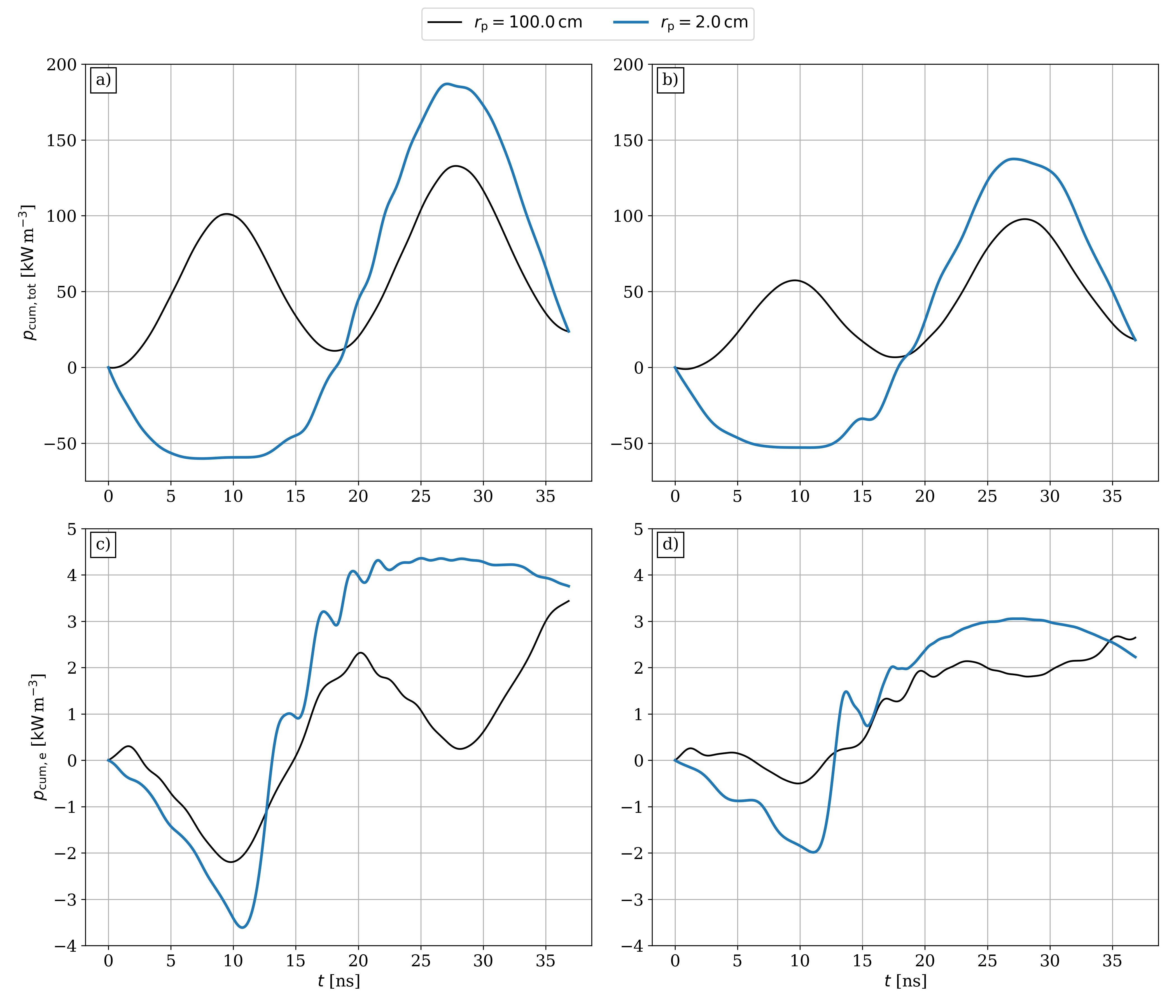}
    \caption{Comparison of the total cumulative power density (first row) and the cumulative electron power density (second row) for \( L_{\mathrm{g}} = 3.0\ \rm{cm} \) (left column) and \( L_{\mathrm{g}} = 6.0\ \rm{cm} \) (right column). Discharge parameters: $V_{0} = 500\,\rm{V},\ f=27.12\,\rm{MHz},\ p = 1.0\,\rm{Pa},\ 100\, \% \,\rm{Ar}$}
    \label{fig:CummPowerP}
\end{figure}

In the following analysis, the cumulative power densities are again shown for the most asymmetric configuration (\( r_{\rm{p}} = 2.0\,\mathrm{cm} \)), but with a reduced driving voltage such that the total input power density matches that of the quasi-symmetric reference case (\( r_{\rm{p}} = 100.0\,\mathrm{cm} \)). This approach is similar to the comparison presented in Figure~\ref{fig:DenPowEqualized}. The total cumulative power densities shown in Figure~\ref{fig:CummPowerP}(a) and (b) display the characteristic time-resolved behavior previously seen in Figure~\ref{fig:CummPower} (a) and (b), including the distinct step-like increases associated with sheath expansion phases. Importantly, both configurations dissipate the same amount of power by the end of the RF period, confirming the validity of the comparison. Figure~\ref{fig:CummPowerP}(c) shows that the cumulative electron power density in the asymmetric configuration still exceeds that of the symmetric case for most of the RF cycle. This indicates that the enhanced energy transfer to the electron population is primarily a geometric effect rather than a result of increased power input. Between \(t \approx 10 - 20\,\mathrm{ns} \), a pronounced rise in cumulative power is observed due to sheath expansion at the powered electrode, again surpassing the symmetric configuration. However, the peaks associated with individual electron beam events are less distinct, and the cumulative profile appear smoother. This smoothing is attributed to the reduced electric field strength resulting from the lower driving voltage, which weakens nonlinear excitation while preserving the underlying electron dynamics. In Figure~\ref{fig:CummPowerP}(d), the temporal evolution of the blue curve is similar to that in Figure~\ref{fig:CummPowerP}(c), but with reduced minima and maxima, whereas the black curve exhibits a different temporal evolution. In particular, the improved energy transfer in the symmetric discharge is linked to constructive interference between the two sheaths, which becomes more effective at larger gap distances. The comparison between the two discharge configurations demonstrates that, while asymmetric discharges rely on abrupt sheath expansion to deposit power, quasi-symmetric discharges rely on the constructive interplay of both sheaths at larger electrode gap separations.

To gain further insight into these mechanisms, the following investigations examines the spatio-temporal electron dynamics in detail. Two representative configurations are analyzed: a strongly asymmetric case with \( r_{\rm{p}} = 2.0\,\mathrm{cm} \), and a more symmetric case with \( r_{\rm{p}} = 10.0\,\mathrm{cm} \). This comparison enables a closer examination of the transition from localized, beam-driven power deposition to more distributed, symmetric power absorption. For reference, the electron dynamics in a fully symmetric configuration under similar discharge conditions have been previously discussed in \cite{Noesges2023}.

 \begin{figure}[h!]
    \centering
    \includegraphics[width=\textwidth]{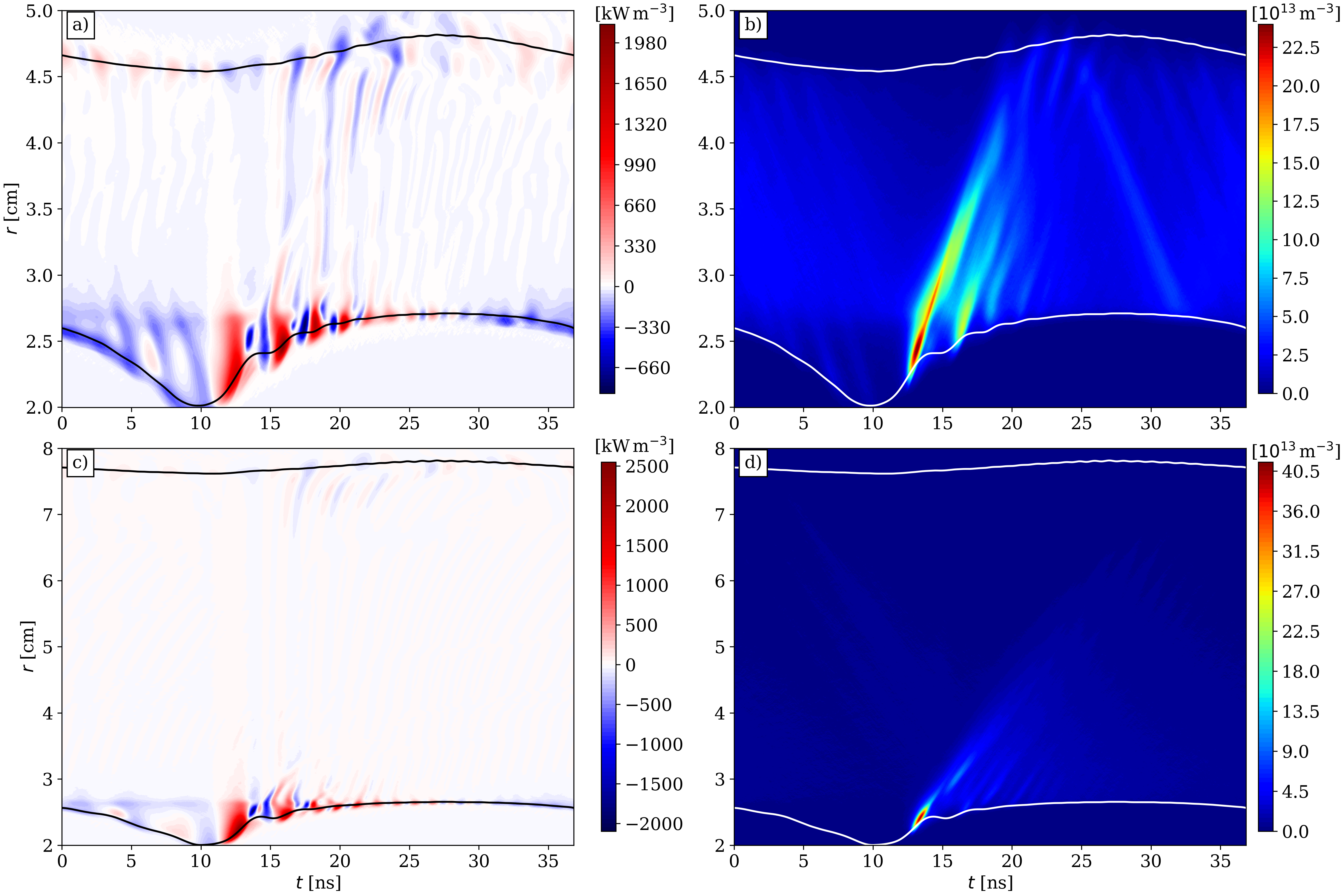}
     \caption{Comparison of the power density ((a) and (c)) and fast electrons moving upwards ((b) and (d)) for \( L_{\mathrm{g}} = 3.0\ \rm{cm} \) (first row) and \( L_{\mathrm{g}} = 6.0\ \rm{cm} \) (second row). Discharge parameters: $r_{\rm{p}}\,=\,2.0\,\rm{cm},\ V_{0} = 500\,\rm{V},\ f=27.12\,\rm{MHz},\ p = 1.0\,\rm{Pa},\ 100\, \% \,\rm{Ar}$}
     \label{fig:XTplots002}
\end{figure}

Figure~\ref{fig:XTplots002} displays the spatio-temporal distribution of the electron power density and the fast electron density (electrons with energies above the ionization threshold) for a small inner radius of \( r_{\rm{p}} = 2.0\,\rm{cm} \) and two different gap sizes: \( L_{\mathrm{g}} = 3.0\,\rm{cm} \) and \( L_{\mathrm{g}} = 6.0\,\rm{cm} \). In Figure~\ref{fig:XTplots002}(a), the electron power density is shown for a small gap size (\( L_{\mathrm{g}} = 3.0\,\rm{cm} \)), with the black line indicating the position of the plasma sheath edge \cite{Brinkmann2009,Brinkmann2015,Klich2022}. Due to the strong asymmetry, power gain and loss are predominantly concentrated near the powered electrode located at \( r = r_{\rm{p}} \). During the expanding sheath phase (\( t \approx 10 - 27\,\mathrm{ns} \)), alternating red and blue regions are clearly visible, representing stepwise electron power gain. This behavior is consistent with the cumulative electron power density observed in Figure~\ref{fig:CummPower}(c). These structures correspond to the acceleration of electrons into multiple beams as they interact with the expanding sheath. These nonlinear dynamics have been well studied in previous works \cite{Mussenbrock2006,Mussenbrock2008,Wilczek2015,Wilczek2016,Wilczek2018,Wilczek2020,Berger2018,Sharma2013,Sharma2014}. At the opposing grounded electrode (\( r = 5.0 \,\rm{cm} \)), blue regions appear during the collapsing sheath phase, caused by electron deceleration. However, no significant power gain is observed at the grounded electrode due to weak sheath modulation in this region. In Figure~\ref{fig:XTplots002}(b), the density of fast electrons (\( \mathcal{E} \geq 15.7\,\mathrm{eV} \)) is shown to further investigate the electron dynamics. During the expanding sheath phase, multiple electron beams are accelerated, forming distinct structures that propagate through the plasma bulk, consistent with the features observed in Figure~\ref{fig:XTplots002}(a). The sheath expands rapidly, forming the first and strongest electron beam, with subsequent beams reducing in intensity over time. Upon reaching the grounded electrode, these beams are partially reflected, resulting in noticeable electron deceleration near the sheath. This reflection enhances the confinement of fast electrons, increasing their contribution to ionization and leading to a relatively high electron density in the discharge, as observed in Figure~\ref{fig:DenPow}(a). 

For the larger gap size (\( L_{\mathrm{g}} = 6.0\,\rm{cm} \)), shown in Figure~\ref{fig:XTplots002}(c), the alternating electron power gain and loss structures near the powered electrode are less pronounced. Instead, the first electron power gain maximum is much more pronounced compared to Figure~\ref{fig:XTplots002}(a). In Figure~\ref{fig:CummPower}(d), the increase in the cumulative electron power density (blue curve) around \( t \approx  12\,\mathrm{ns} \) is linked to this strong initial power gain. At the grounded electrode (\( r = 8.0 \,\rm{cm} \)), almost no red or blue regions are observed, indicating negligible sheath modulation and minimal electron beam interaction in this region due to the larger gap distance. In Figure~\ref{fig:XTplots002}(d), the density of fast electrons reveals a more concentrated acceleration of electrons near the powered electrode. A highly dense beam is accelerated around \( t \approx  11 - 13\,\mathrm{ns} \), corresponding to the first power gain maximum. However, this electron beam propagates a shorter distance into the plasma bulk. The lack of significant interaction with the opposing sheath limits their confinement, resulting in a strongly asymmetric electron density profile (c.f. Figure~\ref{fig:AveDensitiesAr1Pa}(a)). This is because most ionization events occur close to the powered electrode. In conclusion, for asymmetric discharges with a small inner radius, electron power gain and loss are predominantly located at the powered electrode. While the grounded electrode exhibits weak sheath modulation and minimal power gain, its interaction with fast electron beams significantly influences overall electron dynamics. The interactions in strongly asymmetric discharges depend on the gap size and affect global discharge parameters, such as the electron density and ionization rate.

  \begin{figure}[h!]
    \centering
    \includegraphics[width=\textwidth]{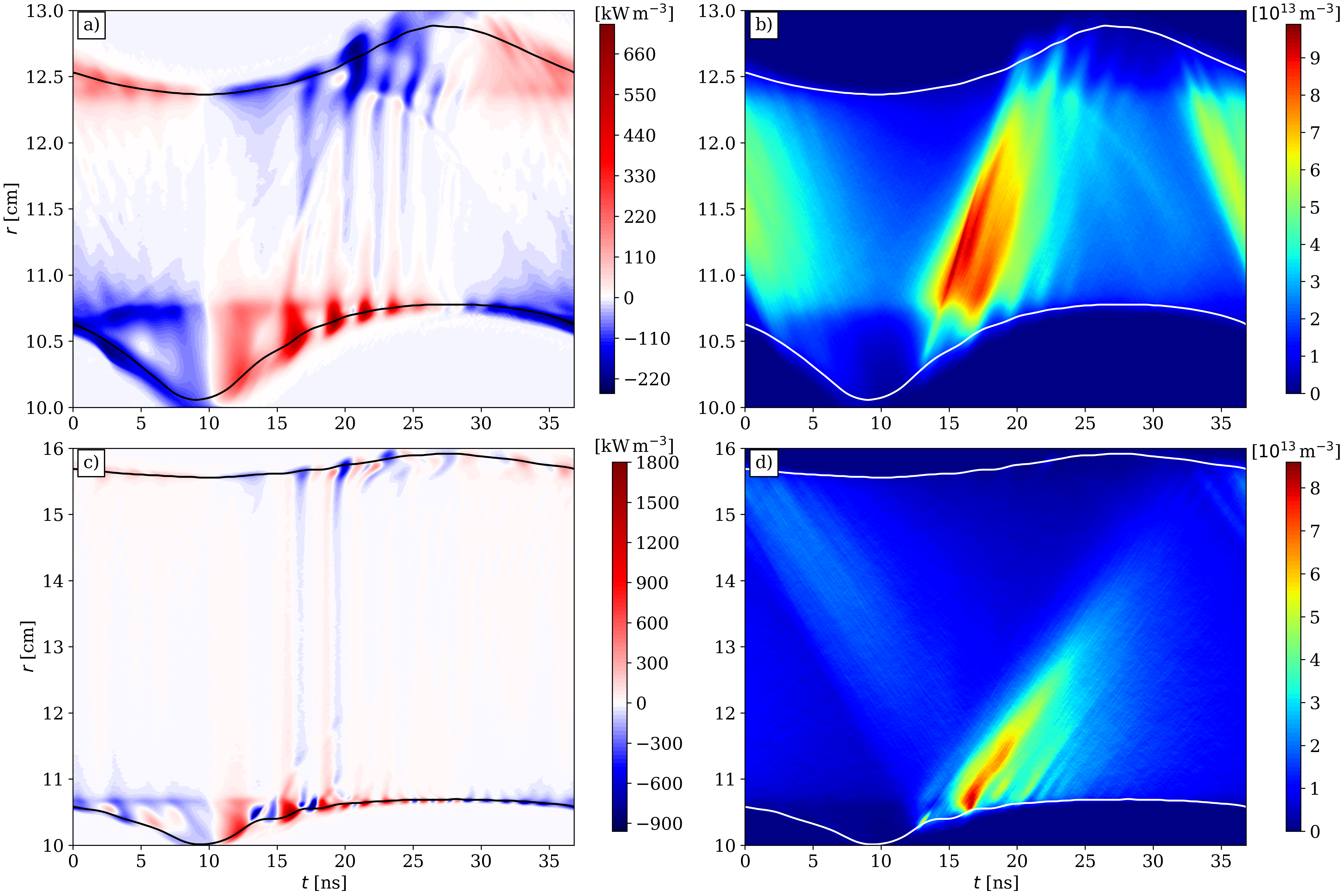}
     \caption{Comparison of the power density ((a) and (c)) and fast electrons moving upwards ((c) and (d))  for \( L_{\mathrm{g}} = 3.0\, \rm{cm} \) (first row) and \( L_{\mathrm{g}} = 6.0\, \rm{cm} \) (second row). Discharge parameters: $r_{\rm{p}}\,=\,10.0\,\rm{m},\ V_{0} = 500\,\rm{V},\ f=27.12\,\rm{MHz},\ p = 1.0\,\rm{Pa},\ 100\, \% \,\rm{Ar}$}
     \label{fig:XTplots010}
 \end{figure}

Figure~\ref{fig:XTplots010} displays the spatio-temporal plots of the electron power density and fast electron density for an inner radius of \( r_{\rm{p}} = 10.0\,\rm{cm} \) and two different gap sizes (\( L_{\mathrm{g}} = 3.0 \,\rm{cm} \), \( L_{\mathrm{g}} = 6.0 \,\rm{cm} \)). Due to the larger inner radius, the asymmetry in this discharge is reduced compared to Figure~\ref{fig:XTplots002} (c.f. Table \ref{AsymmetryFactors}). In Figure~\ref{fig:XTplots010}(a), the electron power density over one RF period and across the gap distance is shown. Electrons gain and lose power at both electrodes, but the effect is slightly stronger at the powered electrode located at \( r = r_{\rm{p}} \). The absolute values of the power gain and loss during the expanding sheath phase (\( t \approx 10 - 27\,\mathrm{ns} \)) are reduced compared to Figure~\ref{fig:XTplots002}(a). At the opposing electrode, the sheath collapses simultaneously, and blue regions indicate significant electron power loss. These structures are non-uniform and are influenced by electron beams reaching the sheath. Additionally, the electron power gain during the expanding sheath phase at the grounded electrode (\( r = 5.0 \,\rm{cm} \)) is weaker and more uniform compared to Figure~\ref{fig:XTplots002}(a). In Figure~\ref{fig:XTplots010}(b), the density of fast electrons (\( \mathcal{E} \geq 15.7\,\mathrm{eV} \)) is displayed, investigating the dynamics of electrons critical for ionization processes. Several electron beams, indistinguishable from each other, are accelerated during the expanding sheath phase. These electron beams interact with the collapsing sheath, where, unlike in Figure~\ref{fig:XTplots002}(b), no significant reflection is observed. Instead, the fast electrons hitting the collapsing sheath are primarily decelerated. In this more symmetric discharge configuration, the opposing sheath is stronger modulated, resulting in alternating electron acceleration and deceleration during the expanding and collapsing phases. Although both sheaths contribute to electron acceleration, the time- and space-averaged plasma density is lower than in the highly asymmetric case with \( r_{\rm{p}} = 2.0\,\rm{cm} \) (c.f. Figure~\ref{fig:DenPow}(a)). This indicates that electron confinement and energy transfer are more efficient in strongly asymmetric discharges at smaller gap sizes. 

In Figure~\ref{fig:XTplots010}(c), the spatio-temporal evolution of the electron power density is shown for a larger gap size of \( L_{\mathrm{g}} = 6.0 \,\rm{cm} \). With the increased gap size, the asymmetry in the discharge is enhanced due to the larger outer radius, and nonlinear effects become more pronounced compared to \( L_{\mathrm{g}} = 3.0 \,\rm{cm} \). During the expanding sheath phase, alternating red and blue regions representing electron power gain and loss are more pronounced compared to Figure~\ref{fig:XTplots010}(a). This observation is consistent with the cumulative power density findings (c.f. Figure~\ref{fig:CummPower}). At the grounded electrode (\( r = 8.0 \,\rm{cm} \)), when electrons reach the collapsing sheath (\( t \approx 10 - 27\,\mathrm{ns} \)), they are decelerated non-uniformly due to the nonlinearity of the system. Figure~\ref{fig:XTplots010}(d) shows that the increased gap size allows most of the multiple electron beams to hit the expanding sheath phase and be partially accelerated back into the bulk plasma, resulting in improved electron confinement. This interaction between electron beams and the expanding sheath phase at the grounded electrode, typically observed in fully symmetric discharges~\cite{Noesges2023,Liu2011,Liu2012,Liu2012b,Liu2013}, leads to enhanced electron confinement and explains the higher electron density observed in these cases. In contrast, Figure~\ref{fig:XTplots002}(d) shows that, in strongly asymmetric discharges, the electron beams do not reach the opposing electrode. Together with the weak sheath modulation at the grounded electrode, this results in poor electron confinement and, consequently, lower electron densities. In summary, the results indicate that larger gap sizes and increased symmetry improve electron confinement by facilitating the interaction between electron beams and the expanding sheath.

 \begin{figure}[h!]
    \centering
    \includegraphics[width=\textwidth]{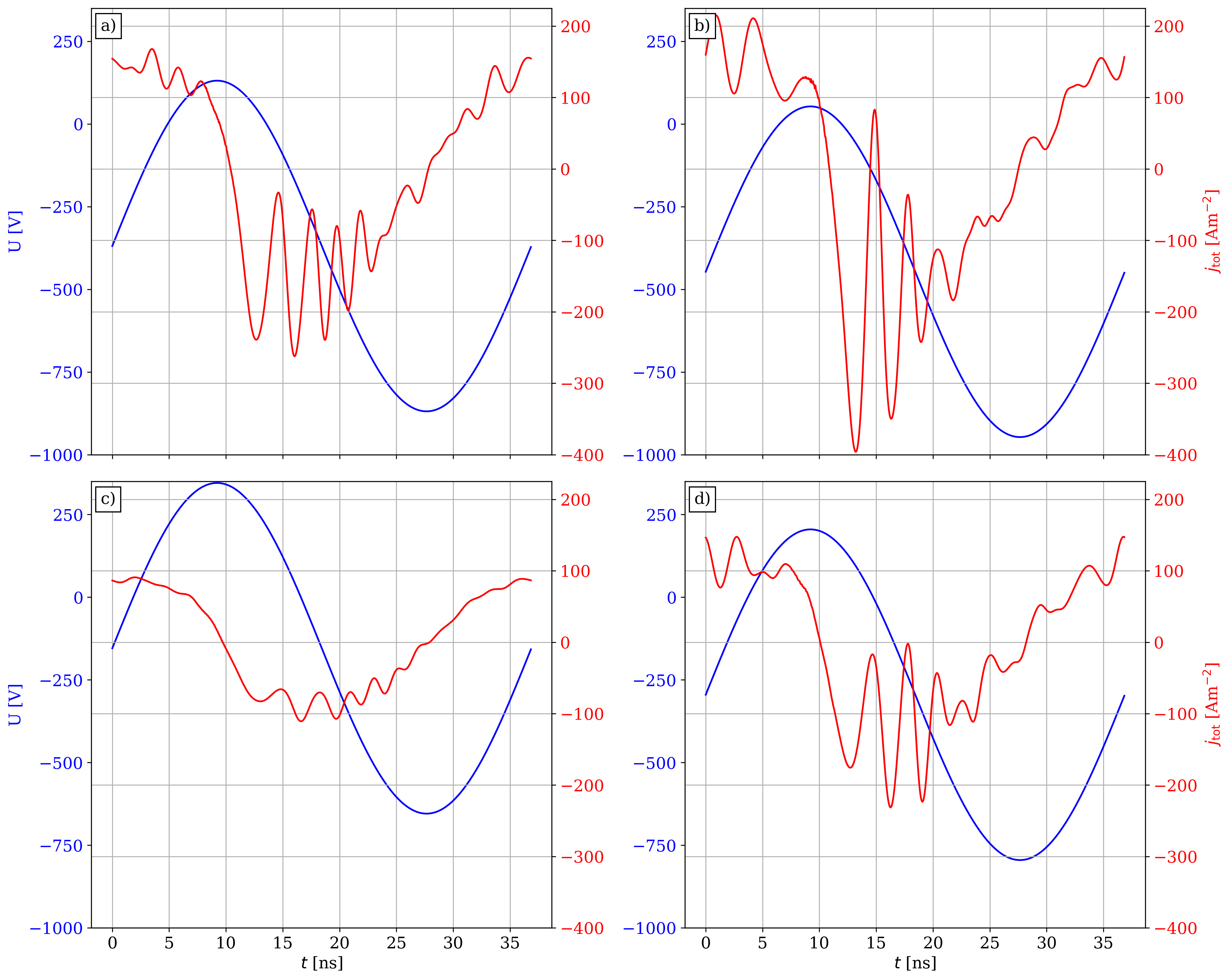}
     \caption{Comparison of the voltage and current density characteristics over time for \( L_{\mathrm{g}} = 3.0\, \rm{cm} \) (left column) and \( L_{\mathrm{g}} = 6.0\, \rm{cm} \) (right column). The first row displays a small inner radius $r_{\rm{p}}\,=\,2.0\,\rm{cm}$ (a and b) and the second row a large inner radius $r_{\rm{p}}\,=\,10.0\,\rm{cm}$ (c and d). Discharge parameters: $V_{0} = 500\,\rm{V},\ f=27.12\,\rm{MHz},\ p = 1.0\,\rm{Pa},\ 100\, \% \,\rm{Ar}$}
    \label{fig:CurrentVoltage}
\end{figure}

Figure~\ref{fig:CurrentVoltage} shows the time evolution of the voltage \( U \) (blue curve) and the total current density \( j_{\mathrm{tot}} \) (red curve) at the powered electrode over one RF period for \( L_{\mathrm{g}} = 3.0\,\mathrm{cm} \) (left column) and \( L_{\mathrm{g}} = 6.0\,\mathrm{cm} \) (right column). The first row presents results for an inner radius of \( r_{\rm{p}} = 2.0\,\mathrm{cm} \), while the second row corresponds to \( r_{\rm{p}} = 10.0\,\mathrm{cm} \). The blue curves shows the applied sinusoidal voltage waveform with an amplitude of \( 500\,\mathrm{V} \), shifted toward negative values due to the presence of a DC self-bias. The red curves depict the total current density \( j_{\mathrm{tot}} \), which exhibits complex behavior due to the nonlinear dynamics of the system. In all cases, the current density exhibits a phase shift relative to the applied voltage, indicating that the discharge impedance approximates that of a capacitor. In Figure~\ref{fig:CurrentVoltage}(a), the DC self-bias is \( V_{\rm{DC}} = 370\,\mathrm{V} \), which causes the applied voltage to remain predominantly negative throughout the RF cycle. The current density initially remains positive but drops rapidly as the voltage decreases. During the sheath expansion phase at the powered electrode (\( t \approx 10 - 27\,\mathrm{ns} \)), strong oscillations emerge due to the excitation of the plasma series resonance. These oscillations gradually dampen over time, and the current density returns to positive values during the sheath expansion at the opposing electrode, closely following the sinusoidal voltage waveform. In Figure~\ref{fig:CurrentVoltage}(b), the DC self-bias reaches its highest value of \( V_{\rm{DC}} = 450\,\mathrm{V} \), resulting from the pronounced asymmetry in the system (\( \epsilon = 16.0 \)). The current density exhibits oscillatory behavior similar to that in Figure~\ref{fig:CurrentVoltage}(a), but with increased damping. This damping is attributed to the larger gap size and therefore, higher plasma density, which modify the resonance conditions. A prominent peak in the current density is observed at \( t \approx 12 \) ns, with a significantly higher peak-to-peak value compared to \( L_{\mathrm{g}} = 3.0\,\mathrm{cm} \). This peak correlates with the formation of a dense electron beam during the sheath expansion at the powered electrode, as shown in Figure~\ref{fig:XTplots002}(d). These dynamics highlight the strong coupling between electron beams and global discharge parameters such as the current waveform. 

Figure~\ref{fig:CurrentVoltage}(c) displays the current and voltage characteristics for a larger inner radius and a small gap distance. Due to the reduced asymmetry (\( \epsilon = 1.69 \)), the DC self-bias is the lowest among the subfigures at \( V_{\rm{DC}} = 150\,\mathrm{V} \), and the modulation of the current density is significantly weaker. Four distinct current density peaks appear during the sheath expansion phase (\( t \approx 10 - 27\,\mathrm{ns} \)), similar to those observed in Figure~\ref{fig:CurrentVoltage}(a), but with significantly lower amplitude. This reduction indicates that the system's nonlinearity is less strongly excited, consistent with the smoother excitation behavior and less distinct electron beams shown in Figure~\ref{fig:XTplots010}(b). Increasing the gap distance, as shown in Figure~\ref{fig:CurrentVoltage}(d), enhances the discharge asymmetry, leading to stronger modulation of the current density. However, the higher plasma density associated with the larger gap results in increased damping of these oscillations. In summary, Figure \ref{fig:CurrentVoltage} illustrates the influence of gap size and inner radius on the electrode voltage and the resulting total current density. The observed plasma series resonance, damping behavior, and beam dynamics underscore the crucial role of geometrical asymmetry in shaping discharge characteristics.

\section{Conclusions and Outlook}

In this work, we have investigated the transition from symmetric to asymmetric capacitively coupled radio-frequency  discharges using a fully kinetic 1d3v electrostatic Particle-in-Cell/Monte Carlo collision simulation in spherical geometry. By systematically varying the inner electrode radius and the electrode gap distance, we explored how geometrical asymmetry influences key plasma parameters, including electron density, power absorption, electron dynamics, and current characteristics. Our results show that increasing asymmetry—achieved by reducing the inner radius—leads to pronounced nonlinear effects, most notably the excitation of the plasma series resonance and the formation of high-energy electron beams. These multiple beams play a dominant role in power coupling and electron confinement, particularly at small gap sizes where beam-driven heating becomes the primary energy transfer mechanism. In contrast, larger inner radii, corresponding to nearly symmetric configurations, result in more balanced power absorption across both sheaths, smoother dynamics, and increased damping of high-frequency oscillations. Notably, at small gap sizes, quasi-symmetric discharges produce lower electron densities, whereas at larger gap sizes, the same configurations lead to significantly higher electron densities due to more efficient electron confinement. Time-resolved analyses of the cumulative power and current density reveal that strongly asymmetric discharges exhibit step-like power gain features, indicative of discrete beam events. Conversely, symmetric cases show smoother, continuous power absorption. Spatio-temporal plots of electron power density and fast electron populations further underline the central role of geometric asymmetry in governing the nonlocal and nonlinear electron dynamics and the resulting discharge behavior. These findings underscore the importance of discharge geometry as a design parameter for controlling plasma properties in CCRF systems. In particular, strong geometric asymmetry can be harnessed to enhance beam-driven power absorption and ionization efficiency, which is beneficial for applications requiring localized or high-density plasmas. In future work, we aim to extend this analysis using a nonlinear global model based on an equivalent circuit representation of the plasma to bridge kinetic and reduced-order modeling approaches. Additionally, the role of instantaneous power absorption in electronegative discharges will be explored to assess the generalizability of these mechanisms across different plasma chemistries.

%\begin{acknowledgments}
%We wish to acknowledge the support of the author community in using
%REV\TeX{}, offering suggestions and encouragement, testing new versions,
%\dots.
%\end{acknowledgments}

\section*{ORCIDs}
\noindent
K. Noesges: \url{https://orcid.org/0000-0002-1579-3301} \\
T. Mussenbrock: \url{https://orcid.org/0000-0001-6445-4990}

\bibliography{bib}% Produces the bibliography via BibTeX.

\end{document}